\documentclass[runningheads]{llncs}
\usepackage[T1]{fontenc}
\usepackage{graphicx}
\usepackage{epsfig}
\usepackage{amsmath}
\usepackage{amssymb}
\usepackage{booktabs}
\usepackage{bbm}
\usepackage{enumitem}
\usepackage{algorithm}
\usepackage[noend]{algorithmic}
\usepackage{caption}
\usepackage{subcaption}
\usepackage{xcolor}
\usepackage[pagebackref=false,breaklinks=true,letterpaper=true,colorlinks,bookmarks=false]{hyperref}

\DeclareMathOperator*{\argmin}{arg\,min}
\DeclareMathOperator{\tr}{tr}

\begin{document}
\title{SMRD: SURE-based Robust MRI Reconstruction with Diffusion Models}

\author{Batu Ozturkler\inst{1}, Chao Liu\inst{2}, Benjamin Eckart\inst{2}, Morteza Mardani\inst{2}, Jiaming Song\inst{2}, Jan Kautz\inst{2}}

 \authorrunning{B. Ozturkler et al.}
\institute{Stanford University, Stanford, CA 94305, USA\\ 
\email{ozt@stanford.edu} \and NVIDIA Corporation, Santa Clara, CA 95051, USA \\
\email{\{chaoliu,	
beckart,	
mmardani,
jkautz\}@nvidia.com} \\
\email{jiaming.tsong@gmail.com}}
\maketitle              
\begin{abstract}
 Diffusion models have recently gained popularity for accelerated MRI reconstruction due to their high sample quality. They can effectively serve as rich data priors while incorporating the forward model flexibly at inference time, and they have been shown to be more robust than unrolled methods under distribution shifts. However, diffusion models require careful tuning of inference hyperparameters on a validation set and are still sensitive to distribution shifts during testing. To address these challenges, we introduce SURE-based MRI Reconstruction with Diffusion models (SMRD), a method that performs test-time hyperparameter tuning to enhance robustness during testing. SMRD uses Stein's Unbiased Risk Estimator (SURE) to estimate the mean squared error of the reconstruction during testing. SURE is then used to automatically tune the inference hyperparameters and to set an early stopping criterion without the need for validation tuning. To the best of our knowledge, SMRD is the first to incorporate SURE into the sampling stage of diffusion models for automatic hyperparameter selection. SMRD outperforms diffusion model baselines on various measurement noise levels, acceleration factors, and anatomies, achieving a PSNR improvement of up to 6 dB under measurement noise. The code is publicly available at \url{https://github.com/NVlabs/SMRD}.
\end{abstract}

\section{Introduction}
Magnetic Resonance Imaging (MRI) is a widely applied imaging technique for medical diagnosis since it is non-invasive and able to generate high-quality clinical images without exposing the subject to radiation. The imaging speed is of vital importance for MRI \cite{SparseMRI07} in that long imaging limits the spatial and temporal resolution and induces reconstruction artifacts such as subject motion during the imaging process. One possible way to speed up the imaging process is to use multiple receiver coils, and reduce the amount of captured data by subsampling the k-space and exploiting the redundancy in the measurements \cite{sense,SparseMRI07}.
 
\begin{figure}[!t]
    \centering
    \includegraphics[width=1.0\textwidth]{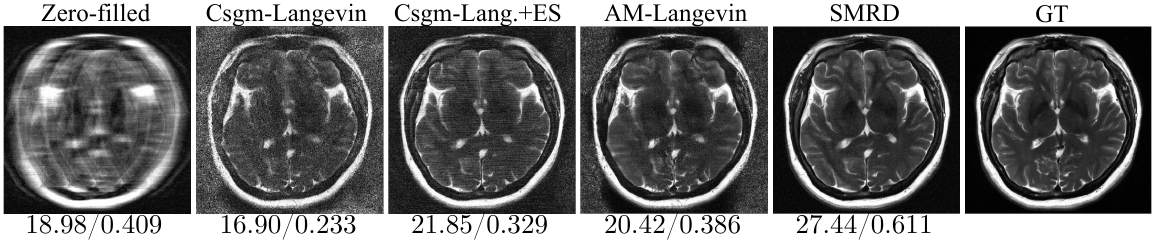}
    \caption{
    Reconstruction of a brain slice from the FastMRI dataset with acceleration rate $R=8$, and measurement noise level $\sigma=0.005$. SMRD is robust to the measurement noise.
     Metrics are reported as \textit{PSNR$/$SSIM}.
    }
    \label{fig:brain}
\end{figure}

In recent years, Deep-Learning (DL) based methods have shown great potential as a data-driven approach in achieving faster imaging and better reconstruction quality. 
DL-based methods can be categorized into two families: unrolled methods that alternate between measurement consistency and a regularization step based on a feed-forward network \cite{Yang16NIPS,hammernik2018learning,ozturkler2022gleam}; conditional generative models that use measurements as guidance during the generative process \cite{mardani2017deep,jalal2021robust}. 
Compared with unrolled methods, generative approaches have recently been shown to be more robust when test samples are out of the training distribution due to their stronger ability to learn the data manifold \cite{chung2022improving,jalal2021robust}. Among generative approaches,  
diffusion models have recently achieved the state of the art performance  \cite{Chung21ScoreMRI,chung2022improving,chung2022solving,xie2022measurement,dar2022adaptive,song2022solving,levac2022accelerated,ozturkler2023regularization}.

However, measurements in MRI 
are often noisy due to the imaging hardware and thermal fluctuations of the subject \cite{snr-mri}. As a result, DL-based methods fail dramatically when a distribution shift due to noise or other scanning parameters occurs during training and testing \cite{antun,snrtraintest}. 
Although diffusion models were shown to be robust against distribution shifts in noisy inverse problems  \cite{song2023pseudoinverseguided,chung2023diffusion} and MRI reconstruction, the hyperparameters that balance the measurement consistency and prior are tuned manually during validation where ground truth is available \cite{jalal2021robust}. These hyperparameters may not generalize well to test settings as shown in Fig.\ref{fig:brain}, and ground truth data may not be available for validation.

In this paper, we propose a framework with diffusion models for MRI reconstruction that is robust to measurement noise and distribution shifts. To achieve robustness, we perform test-time tuning when ground truth data is not available at test time, using Stein’s Unbiased Risk Estimator (SURE) \cite{Stein1981EstimationOT} as a surrogate loss function for the true mean-squared error (MSE). SURE is used to tune the weight of the balance between measurement consistency and the learned prior for each diffusion step such that the measurement consistency adapts to the measurement noise level during the inference process. SURE is then used to perform early stopping to prevent overfitting to measurement noise at inference. 

We evaluate our framework on FastMRI \cite{zbontar2018fastmri} and Mridata \cite{ong2018mridata}, and show that it achieves state-of-the-art performance across different noise levels, acceleration rates and anatomies without fine-tuning the pre-trained network or performing validation tuning, with no access to the target distribution. In summary, our contributions are three-fold: 

\begin{itemize}[leftmargin=*]
    \item We propose a test-time hyperparameter tuning algorithm that boosts the robustness of the pre-trained diffusion models against distribution shifts;
    \item We propose to use SURE as a surrogate loss function for MSE and incorporate it into the sampling stage for test-time tuning and early stopping without access to ground truth data from the target distribution; 
    \item SMRD achieves state-of-the-art performance across different noise levels, acceleration rates, and anatomies.
\end{itemize}

\section{Related Work}
\textbf{SURE for MRI Reconstruction.} 
SURE has been used for tuning parameters of compressed sensing \cite{sidsure}, as well as for unsupervised training of DL-based methods for MRI reconstruction \cite{aggarwal2022ensure}. In addition, two concurrent works propose to incorporate SURE as a loss function for diffusion model training for MRI reconstruction \cite{kawar2023gsure,aali2023solving}. To the best of our knowledge, SURE has not yet been applied to the sampling stage of diffusion models in MRI reconstruction or in another domain.\\
\textbf{Adaptation in MRI Reconstruction.} Several proposals have been made for adaptation to a target distribution in MRI reconstruction using self-supervised losses for unrolled models \cite{yaman2022zeroshot,darestani2022test}. Later, \cite{arvinte2022single} proposed single-shot adaptation for test-time tuning of diffusion models by performing grid search over hyperparameters with a ground truth from the target distribution. However, this search is computationally costly, and the assumption of access to samples from the target distribution is limiting for imaging cases where ground truth is not available. 

\section{Method}

\subsection{Accelerated MRI Reconstruction using Diffusion Models}

 The sensing model for accelerated MRI can be expressed as
 \begin{equation}
     y =  \Omega FSx + \nu
 \end{equation}
  where $y$ is the measurements in the Fourier domain (k-space), $x$ is the real image, $S$ are coil sensitivity maps, $F$ is the Fourier transform, $\Omega$ is the undersampling mask, $\nu$ is additive noise, and $A =\Omega FS$ denotes the forward model.
 
Diffusion models are a recent class of generative models showing remarkable sample fidelity for computer vision tasks \cite{ho2020denoising}. A popular class of diffusion models is score matching with Langevin dynamics \cite{song2019generative}. Given \emph{i.i.d.} training samples from a high-dimensional data distribution (denoted as $p(x)$), diffusion models can estimate the scores of the noise-perturbed data distributions. First, the data distribution is perturbed with Gaussian noise of different intensities with standard deviation $\beta_t$ for various timesteps $t$, such that $p_{\beta_t}(\Tilde{x}|x) = \mathcal{N}(\Tilde{x}|x,\beta_t^2I)$ \cite{song2019generative}, leading to a series of perturbed data distributions $p(x_t)$. Then, the score function $\nabla_{x_t} \log p(x_t)$ can be estimated by training a joint neural network, denoted as $f(x_t;t)$, via denoising score matching~\cite{vincent2011connection}. 
After training the score function, annealed Langevin Dynamics can be used to generate new samples \cite{song2020improved}. Starting from a noise distribution $x_0 \sim \mathcal{N}(0,I)$,  annealed Langevin Dynamics is run for $T$ steps
\begin{equation}
    x_{t+1} = x_t + \eta_t\nabla_{x_t} \log p(x_t) + \sqrt{2\eta_t}\zeta_t
    \label{eq:csgm}
\end{equation}
 where $\eta_t$ is a sampling hyperparameter, and $\zeta_t \sim \mathcal{N}(0,I)$. 
 For MRI reconstruction, measurement consistency can be incorporated via sampling from the posterior distribution $p(x_t|y)$ \cite{jalal2021robust}:  
\begin{equation}
    x_{t+1} = x_t + \eta_t\nabla_{x_t} \log p(x_t|y) +  \sqrt{2\eta_t}\zeta_t
    \label{eq:csgm}
\end{equation}
The form of $\nabla_{x_t} \log p(x_t|y)$ depends on the specific inference algorithm \cite{Chung21ScoreMRI,chung2022improving}.

\begin{algorithm}[t!]
 \caption{SMRD: Test-time Tuning for Diffusion Models via SURE}
 \label{alg:gr-alg}
 \begin{algorithmic}[1]
 \renewcommand{\algorithmicrequire}{\textbf{Input:}}
 \REQUIRE measurement $y$, forward model $A$, initial $\lambda_0$, window size $w$, learning rate $\alpha$
  \STATE sample $x_0 \sim \mathcal{N}(0,I)$
  \FOR {$t \in 0,...,T-1$}
  \STATE $x^+_t = x_t + \eta_tf(x_t;t) + \sqrt{2\eta_t}\zeta_t$ \\
  \STATE $x_{t+1} = h(x_t,\lambda_t) = (A^HA + \lambda_t I)^{-1}(x_{zf} + \lambda_t x^+_t)$ \\
  \STATE sample $\mu \sim \mathcal{N}(0,I) $
  \STATE $\text{SURE}(t) =
     \|h(x_t,\lambda_t) - x_{\mathrm{zf}}\|^2  \mu^T(h(x_t+\epsilon\mu,\lambda_t)-h(x_t,\lambda_t)) / {N\epsilon} $
   \STATE $\lambda_{t+1} = \lambda_{t} - \alpha \nabla_{\lambda_t}\text{{SURE}}(t)$
   \IF {\textit{mean}$\left( \text{SURE}[t-w:t] \right)$ $>$
   \textit{mean}$\left(\text{SURE}[t-2w:t-w]\right)$}
       \STATE \textbf{break}
   \ENDIF
  \ENDFOR
 \RETURN $x_{t+1}$
 \end{algorithmic} 
 \end{algorithm}

\subsection{Stein's Unbiased Risk Estimator (SURE)}
SURE is a statistical technique which serves as a surrogate for the true mean squared error (MSE) when the ground truth is unknown. Given the ground truth image $x$, the zero-filled image can be formulated as $x_{\mathrm{zf}} = x + z$, where $z$ is the noise due to undersampling. Then, SURE is an unbiased estimator of $\text{MSE}=\| \hat{x} - x \|_2^2$ \cite{Stein1981EstimationOT} and can be calculated as:
\begin{equation}
  \text{SURE} = \|\hat{x} - x_{\mathrm{zf}}\|_2^2 -N\sigma^2 + \sigma^2 \tr(\frac{\partial \hat{x}}{\partial x_{\mathrm{zf}}})
\end{equation}
where $x_{zf}$ is the input of a denoiser, $\hat{x}$ is the prediction of a denoiser, $N$ is the dimensionality of $\hat{x}$. In practical applications, the noise variance $\sigma^2$ is not known a priori. In such cases, it can be assumed that the reconstruction error is not large, and the sample variance between the zero-filled image and the reconstruction can be used to estimate the noise variance, where $\sigma^2 \approx \|\hat{x} - x_{\mathrm{zf}}\|_2^2/N$ \cite{edupuganti2020uncertainty}. Then, SURE can be rewritten as:
\begin{equation}
  \text{SURE} \approx \frac{\|\hat{x} - x_{\mathrm{zf}}\|^2}{N} \tr(\frac{\partial \hat{x}}{\partial x_{\mathrm{zf}}})
\end{equation}

A key assumption behind SURE is that the noise process that relates the zero-filled image to the ground truth is i.i.d. normal, namely $z \sim \mathcal{N}(0,\sigma^2I)$. 
However, this assumption does not always hold in the case of MRI reconstruction due to undersampling in k-space that leads to structured aliasing.
In this case, density compensation can be applied to enforce zero-mean residuals and increase residual normality \cite{edupuganti2020uncertainty}.

\begin{figure}[!t]
    \centering
    \includegraphics[width=1.0\textwidth]{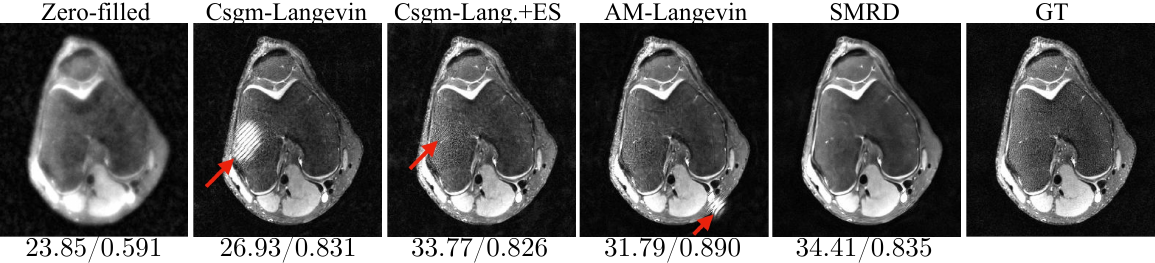}
    \caption{Knee reconstruction with $R=12$.
    Metrics are reported as \textit{PSNR$/$SSIM}.
    }
    \label{fig:knee}
\end{figure}

\subsection{SURE-based MRI Reconstruction with Diffusion Models}
Having access to SURE at test time enables us to monitor it as a proxy for the true MSE. Our goal is to optimize inference hyperparameters in order to minimize SURE. To do this, we first consider the following optimization problem at time-step $t$: 
 \begin{equation}
    \label{eq:forward}
     x_{t+1} = \argmin_{x} \|Ax-y\|_{2}^{2}+\lambda_t \|x-x_t\|_2^{2}
 \end{equation}
 where we introduce a time-dependent regularization parameter $\lambda_t$. The problem in Eq.~\ref{eq:forward} can be solved using alternating minimization (which we call AM-Langevin):
\begin{align}
    x^+_t &= x_t + \eta_tf(x_t;t) + \sqrt{2\eta_t}\zeta_t \\
    x_{t+1} &= \argmin_x \|Ax-y\|_2^2 + \lambda_t\|x-x^+_t\|_2^2
    \label{eq:am-langevin}
\end{align}
Eq. \ref{eq:am-langevin} can be solved using the conjugate gradient (CG) algorithm, with the iterates 
\begin{equation}
\label{eq:xt+1}
    x_{t+1} = h(x_t,\lambda_t) = (A^HA + \lambda_t I)^{-1} (x_{\mathrm{zf}} + \lambda_t x^+_t)
\end{equation}
where $A^H$ is the Hermitian transpose of $A$, $x_{\mathrm{zf}}=A^Hy$ is the zero-filled image, and $h$ denotes the full update including the Langevin Dynamics and CG.   
This allows us to explicitly control the balance between the prior through the score function and the regularization through $\lambda_t$.

\noindent \textbf{Monte-Carlo SURE}. Calculating SURE requires evaluating the trace of the Jacobian $\tr(\frac{\partial x_{t+1}}{\partial x_{\mathrm{zf}}})$, which can be computationally intensive. Thus, we approximate this term using Monte-Carlo SURE \cite{mc-sure}. Given an $N$-dimensional noise vector $\mu$ from $\mathcal{N}(0,I)$ and the perturbation scale $\epsilon$, the approximation is: 
\begin{equation}
    \tr(\frac{\partial x_{t+1}}{\partial x_{\mathrm{zf}}}) \approx \mu^T(h(x_t+\epsilon\mu,\lambda_t)-h(x_t,\lambda_t)) / \epsilon 
\end{equation}

\begin{table*}[t!]
\centering
\caption{
FastMRI brain dataset results. Metrics are reported as \textit{PSNR$/$SSIM}. 
}
\begin{tabular}{l | c c c  | c c c }
\toprule 
$R$ &  \multicolumn{3}{c|}{$R=4$} & \multicolumn{3}{c}{$R=8$}
 \\ $\sigma$ $\left( \times 10^{-3} \right)$ & $\sigma=0$ & $\sigma=2.5$ & $\sigma=5 $  
 & $\sigma=0$ & $\sigma=2.5$ & $\sigma=5 $  \\
\midrule 
    Zero-filled & $27.8 / 0.81$ & $27.1 / 0.63$  & $25.3/ 0.43$  
     &$23.2 / 0.69$  & $23.1 / 0.61$ &$22.7 / 0.43$  \\ 
     Csgm-Langevin & $36.3/0.78$ & $25.1 / 0.36$ & $18.5 / 0.16$  
     & $\mathbf{34.7}/0.79$ & $21.3 / 0.32$ & $15.8/0.15$  \\ 
     Csgm-Lang.+ES & $35.9/0.79$ & $26.1 / 0.40$ & $20.8/0.23$  
     & $32.3/0.69$ & $26.1 / 0.41$ & $20.7/0.25$ \\
     AM-Langevin & $\mathbf{39.7/0.95}$ & $28.7 / 0.53$ & $22.4/0.29$  
     &$ \textbf{34.7} / \mathbf{0.93}$& $25.7 / 0.52$ & $19.0 / 0.30 $ \\
     SMRD & $36.5/0.89$ & $\mathbf{33.5 / 0.78}$ &$ \mathbf{29.4/ 0.59}$ 
     & $32.4 / 0.80$ & $\mathbf{31.4 / 0.75}$ & $\mathbf{28.6}/\mathbf{0.61}$   \\
\bottomrule
\end{tabular}

\label{tbl:comparison_same_anatomy}
\end{table*}
This approximation is typically quite tight for some small value $\epsilon$; see e.g., \cite{edupuganti2020uncertainty}. Then, at time step $t$, SURE is given as 
\begin{equation}
  \text{SURE}(t) = \frac{\|h(x_t,\lambda_t) - x_{\mathrm{zf}}\|^2}{N\epsilon} \mu^T(h(x_t+\epsilon\mu,\lambda_t)-h(x_t,\lambda_t))  
\end{equation}
where $h(x_t,\lambda_t)$ is the prediction which depends on the input $x_{zf}$, shown in Eq. \ref{eq:xt+1}.\\
\noindent\textbf{Tuning $\lambda_t$.} By allowing $\lambda_t$ to be a learnable, time-dependent variable, we can perform test-time tuning (TTT) for  $\lambda_t$ by updating it in the direction that minimizes SURE. 
As both $\text{SURE}(t)$ and $\lambda_t$ are time-dependent, the gradients can be calculated with backpropagation through time (BPTT). In SMRD, for the sake of computation, we apply truncated BPTT, and only consider gradients from the current time step $t$. Then, the $\lambda_t$ update rule is:
\begin{equation}
    \lambda_{t+1} = \lambda_{t} - \alpha \nabla_{\lambda_t}\text{{SURE}}(t)
\end{equation}
where $\alpha$ is the learning rate for $\lambda_t$.\\
\textbf{Early Stopping (ES).} Under measurement noise, it is critical to prevent overfitting to the measurements. We employ early-stopping (ES) by monitoring the moving average of SURE loss with a window size $w$ at test time. Intuitively, we perform early stopping when the SURE loss does not decrease over a certain window. We denote the early-stopping iteration as $T_{ES}$. Our full method is shown in Alg. \ref{alg:gr-alg}.

\section{Experiments}

Experiments were performed with PyTorch on a NVIDIA Tesla V100 GPU \cite{paszke2019pytorch}. For baselines and SMRD, we use the score function from \cite{jalal2021robust} which was trained on a subset of the FastMRI multi-coil brain dataset. We refer the reader to \cite{jalal2021robust} for implementation details regarding the score function. For all AM-Langevin variants, we use 5 CG steps. In SMRD, for tuning $\lambda_t$, we use the Adam optimizer \cite{kingma2017adam} with a learning rate of $\alpha = 0.2$ and $\lambda_0=2$. In the interest of inference speed, we fixed $\lambda_t$ after $t = 500$, as convergence was observed in earlier iterations. For SURE early stopping, we use window size $w=160$. 
For evaluation, we used the multi-coil fastMRI brain dataset \cite{zbontar2018fastmri} and the fully-sampled 3D fast-spin echo multi-coil knee MRI dataset from mridata.org \cite{ong2018mridata} with 1D equispaced undersampling and a 2D Poisson Disc undersampling mask respectively, as in \cite{jalal2021robust}. We used $6$ volumes from the validation split for fastMRI, and $3$ volumes for Mridata where we selected $32$ middle slices from each volume so that both datasets had $96$ test slices in total.\\ 
 \noindent\textbf{Noise Simulation.}~The noise source in MRI acquisition is modeled as additive complex-valued Gaussian noise added to each acquired k-space sample \cite{macovski1996noise}. To simulate measurement noise, we add masked complex-gaussian noise to the masked k-space $\nu \sim \mathcal{N}(0,\sigma)$ with standard deviation $\sigma$, similar to \cite{desai2023noise2recon}.

\begin{table*}[t!]
\centering
\caption{Cross-dataset results with the Mridata knee dataset.
Metrics are reported as \textit{PSNR$/$SSIM}.}
\begin{tabular}{l | c c c  | c c c  }
\toprule 
$R$
 &  \multicolumn{3}{c|}{$R=12$} & \multicolumn{3}{c}{$R=16$}
 \\ 
 $\sigma$ 
 $\left( \times 10^{-3} \right)$ 
 & $\sigma=0$ & $\sigma=2.5$ & $\sigma=5 $
 & $\sigma=0$ & $\sigma=2.5$ & $\sigma=5 $\\
\midrule 
    Zero-filled & $24.5/0.63$ & $24.5 / 0.61$ & $24.1 / 0.54$ 
     & $24.0 / 0.60$ & $24.1 / 0.59$ & $23.9 / 0.55$ \\
     Csgm-Langevin 
     & $31.4/0.82$ & $24.2/0.49$ & $21.7/0.29$ 
     & $31.8/0.79$ & $24.4/0.50$ & $22.1/0.32$ \\ 
     Csgm-Lang.+ES 
     & $34.0/0.82$ & $28.8/0.60$ & $23.6/0.37$
     & $33.6/0.81$ & $28.4/0.60$ & $23.2/0.37$\\
     AM-Langevin 
     & $34.0/\mathbf{0.87}$ & $29.9/0.69$ & $23.5/0.38$
     & $\mathbf{34.3/0.85}$ & $29.3/0.66$ & $22.7/0.37$\\
     SMRD & $\mathbf{35.0}/0.84$ & $\mathbf{33.5/0.82}$ & $\mathbf{29.4/0.67}$
     & $34.1/0.82$ & $\mathbf{32.8/0.80}$ & $\mathbf{29.2/0.67}$\\
\bottomrule
\end{tabular}
\label{tbl:comparison_different_anatomy}
\end{table*}

\begin{table}[t!]
\centering
\caption{Ablation study for different components of our method on the Mridata multi-coil knee dataset where $R = 12$. Metrics are reported as \textit{PSNR}/\textit{SSIM}.\small}
\begin{tabular}{l | c c c c }
\toprule
$\sigma$  ($\times 10^{-3}$) &  \multicolumn{1}{c}{$\sigma = 0$} & \multicolumn{1}{c}{$\sigma = 2.5$}  & \multicolumn{1}{c}{$\sigma = 5$} & \multicolumn{1}{c}{$\sigma = 7.5$}
 \\ 
\midrule 
     AM-Langevin & $34.0/\mathbf{0.87}$ & $29.9/0.69$ & $23.5/0.38$ & $20.5/0.24$ \\ 
     AM-Lang.+TTT & $34.7/0.86$ & $31.0 / 0.75$ & $25.1/0.45$ & $21.5 / 0.29$\\
     AM-Lang.+ES & $\mathbf{35.3}/0.85$ & $32.9/0.80$ &  $28.3/0.61$ & $25.2/0.45$ \\
     AM-Lang.+TTT+ES (SMRD)& $35.0/0.84$ & $\mathbf{33.5/0.82}$ & $\mathbf{29.4/0.67}$ & $\mathbf{26.3/0.52}$  \\
\bottomrule
\end{tabular}

\label{tbl:ablations}
\end{table}

\begin{figure}
    \centering 
    \begin{subfigure}{.35\textwidth}
         \centering
         \includegraphics[width=\textwidth]{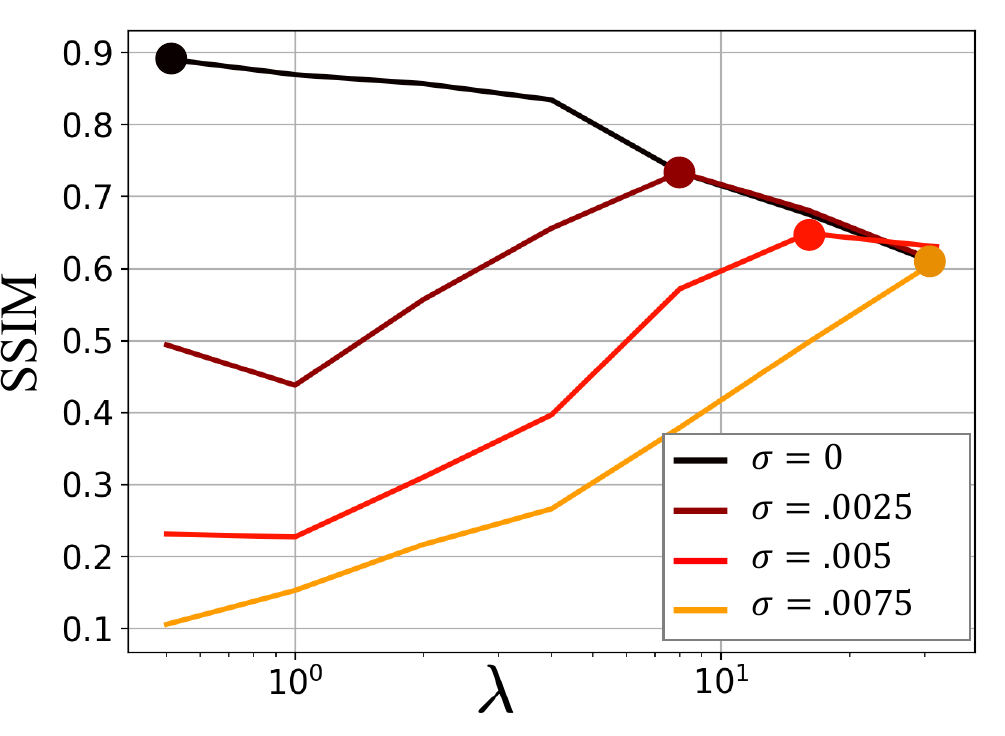}
         \caption{SSIM \textit{vs} $\lambda$}
         \label{fig:ssim_vs_lambda}
    \end{subfigure}
    \begin{subfigure}{.35\textwidth}
        \centering
        \includegraphics[width=\textwidth]{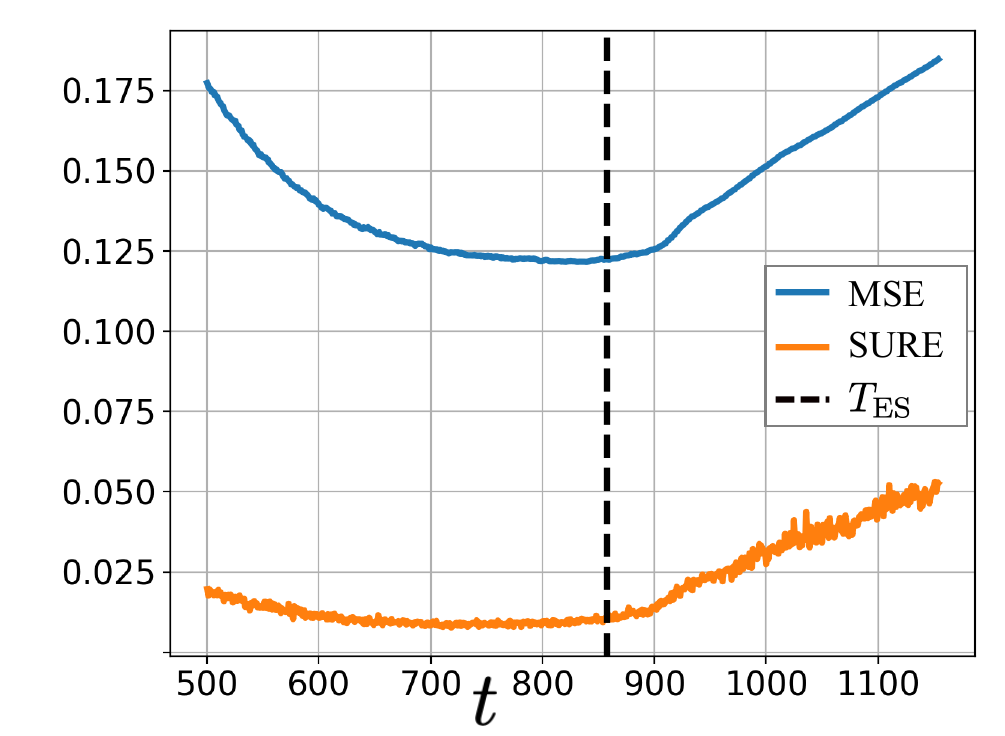}
        \caption{SURE and MSE loss 
        }
        \label{fig:sure_vs_gt}
    \end{subfigure}
    \caption{
    (a)~SSIM on a validation scan (knee) with varying noise levels $\sigma$ and regularization parameter $\lambda$ parameter choices. As $\sigma$ increases, the optimal $\lambda$ value shifts to the right. 
    (b)~SURE and true MSE over sampling iterations. MSE and SURE start increasing at similar $T$ due to overfitting to measurement noise.}
    \label{fig:losses}
\end{figure}

\begin{figure}[!t]
    \centering
    \includegraphics[width=1.0\textwidth,trim={0 2pt 0 0},clip]{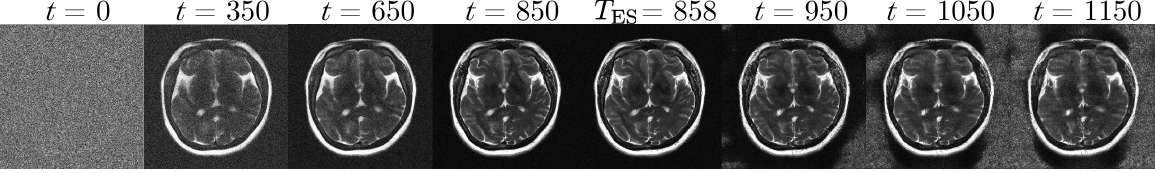}
    \caption{
    Iterations of inference for an example brain slice where $R = 8$, $\sigma = 0.0075$.
    }
    \label{fig:evol}
\end{figure}

\section{Results and Discussion}

 We compare with three baselines:
 (1)~Csgm-Langevin \cite{jalal2021robust}, using the default hyperparameters that were tuned on two validation brain scans at $R = 4$;
 (2)~Csgm-Langevin with early stopping using SURE loss with window size $w=25$;
 (3)~AM-Langevin where $\lambda_t=\lambda_0$ and fixed throughout the inference. 
 For tuning AM-Langevin, we used a brain scan for validation at $R = 4$ in the noiseless case ($\sigma = 0$) similar to Csgm-Langevin, and the optimal value was $\lambda_{0}=2$. 
 
We evaluate the methods across different measurement noise levels, acceleration rates and anatomies using the same pretrained score function from \cite{jalal2021robust}.
Table \ref{tbl:comparison_same_anatomy} shows a comparison of reconstruction methods applied to the FastMRI brain dataset where $R = \{4,8\}$, and $\sigma=\{0,0.0025,0.005\}$. SMRD performs best except for $\sigma=0$, which is the training setting for the score function. Example reconstructions are shown in Fig.\ref{fig:brain} where $R=8$, $\sigma=0.005$. Under measurement noise, reconstructions of baseline methods diverge, where only SMRD can produce an acceptable reconstruction. 
Table \ref{tbl:comparison_different_anatomy} shows a comparison of reconstruction methods in the cross-dataset setup with the Mridata knee dataset where $R = \{12,16\}$. SMRD outperformed baselines across every $R$ and $\sigma$, and is on par with baselines when $\sigma=0$. Fig.\ref{fig:knee} shows example reconstructions where $R=12$, $\sigma=0$. Hallucination artifacts are visible in baselines even with no added measurement noise, whereas SMRD mitigates these artifacts and produces a reconstruction with no hallucinations. Fig.\ref{fig:ssim_vs_lambda} shows SSIM on a knee validation scan with varying noise levels $\sigma$ and $\lambda$. As $\sigma$ increases, the optimal $\lambda$ value increases as well. Thus, hyperparameters tuned with $\sigma=0$ do not generalize well under measurement noise change, illustrating the need for test-time tuning under distribution shift. Fig.\ref{fig:sure_vs_gt} shows true MSE vs SURE for an example brain slice where $\sigma=0.0075$, $R=8$. SURE accurately estimates MSE, and the increase in loss occurs at similar iterations, enabling us to perform early stopping before true MSE increases. 
The evolution of images across iterations for this sample is shown in Fig.\ref{fig:evol}. SMRD accurately captures the correct early stopping point. As a result of early stopping, SMRD mitigates artifacts, and produces a smoother reconstruction.
Table \ref{tbl:ablations} shows the ablation study for different components of SMRD. TTT and ES both improve over AM-Langevin, where SMRD works best for all $\sigma$ while being on par with others on $\sigma=0$.

\section{Conclusion}
We presented SMRD, a SURE-based TTT method for diffusion models in MRI reconstruction. SMRD does not require ground truth data from the target distribution for tuning as it uses SURE for estimating the true MSE. SMRD surpassed baselines across different shifts including anatomy shift, measurement noise change, and acceleration rate change. SMRD could be helpful to improve the safety and robustness of diffusion models for MRI reconstruction used in clinical settings. While we applied SMRD to MRI reconstruction, future work could explore the application of SMRD to other inverse problems and diffusion sampling algorithms and can be used to tune their inference hyperparameters. 

\newpage
\bibliographystyle{splncs04}
\bibliography{main_arxiv}

\newpage

\section{Supplementary Material}

 \subsubsection{Mridata Multi-Coil Knee Dataset.}
 We used the fully-sampled 3D fast-spin echo (FSE) multi-coil knee MRI dataset publicly available on mridata.org \cite{ong2018mridata}. Each 3D volume had a matrix size of $320 \times 320 \times 256$ with $8$ coils. Sensitivity map estimation was performed in SigPy \cite{sigpy} using JSENSE \cite{jsense} with kernel width of 8 for each volume. Fully-sampled references were retrospectively undersampled using a 2D Poisson Disc undersampling mask at $R = \{12,16\}$, where $R$ denotes acceleration factor.  For evaluation, we used 3 volumes, and selected $32$ middle slices from each volume for a total of $96$ test slices.
\subsubsection{FastMRI Brain Multi-Coil Dataset.}
 We used the multi-coil brain MRI dataset from fastMRI \cite{zbontar2018fastmri}. The matrix size for each scan was $384 \times 384$ with $15$ coils. Sensitivity maps were estimated using ESPIRiT with a kernel width of $8$ and a calibration region of $12 \times 12$ \cite{uecker2014espirit}. At acceleration factor $R = 8$, this calibration region is $4\%$ of the autocalibration region. Fully-sampled references were retrospectively undersampled using 1D equispaced undersampling, as in \cite{jalal2021robust}. For evaluation, we select 6 volumes from the validation split, and evaluate all methods on $96$ slices.

\begin{table*}[h]
\centering

\begin{tabular}{l | c c | c c }
\toprule 
Anatomy
 &  \multicolumn{2}{c|}{Brain} & \multicolumn{2}{c}{Knee}
 \\ 
 $R$ 
 & $R=4$ & $R=8$ 
 & $R=12$ & $R=16$\\ 
\midrule 
    Zero-filled & $23.4 / 0.31$ 
       & $22.0 / 0.32$ &  $23.8 / 0.48$ 
     & $23.4 / 0.48$ \\ 
     Csgm-Langevin &  $16.0 / 0.08$ 
      & $14.8 / 0.09$ & $18.7/0.16$ 
     & $19.6/0.20$\\ 
     Csgm-Langevin+ES  & $18.0 / 0.16$ 
      & $17.7 / 0.17$ &  $20.5/0.25$ 
     &  $20.6/0.26$\\
     AM-Langevin &  $18.7 / 0.18$ 
     & $15.9 / 0.19$ &  $20.5/0.24$ 
     &  $19.2/0.23$\\
     SMRD &  $\mathbf{26.3 / 0.44}$
     & $\mathbf{26.4 / 0.49}$ & $\mathbf{26.3/0.52}$ 
     & $\mathbf{26.1/0.52}$\\
\bottomrule
\end{tabular}

\caption{
Reconstruction methods applied to the FastMRI multi-coil brain dataset and the Mridata multi-coil knee dataset at $\sigma=0.0075$. Metrics are reported as \textit{PSNR$/$SSIM}. In high noise regimes, SMRD outperforms all baselines. 
}
\label{tbl:comparison_same_anatomy}
\end{table*}

\begin{figure}[h]
    \centering
    \includegraphics[width=1.0\textwidth,trim={0 2pt 0 0},clip]{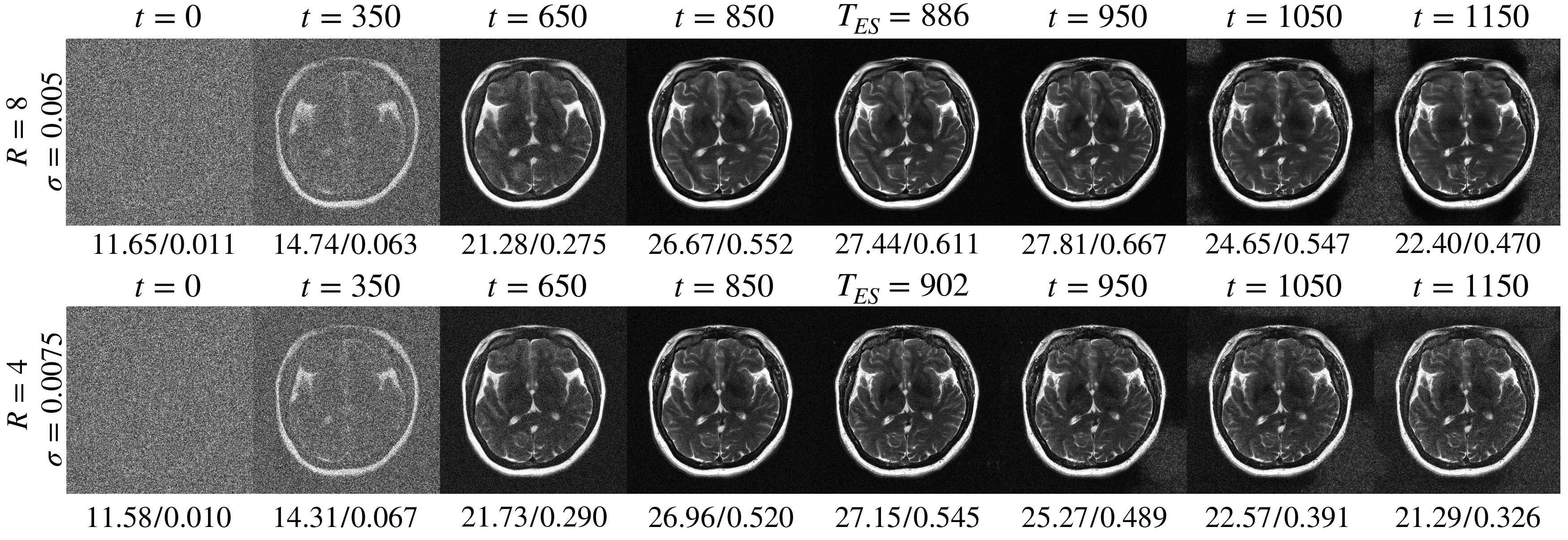}
    \caption{
    Iterations of inference for an example brain slice where $R = 8$, $\sigma = 0.005$ (top), and $R = 4$, $\sigma = 0.0075$ (bottom). Metrics are reported as \textit{PSNR$/$SSIM}. The time step $t$ where divergence is observed for the inference process changes for different $R$ and $\sigma$. As a result, $T_{ES}$ changes. 
    }
    \label{fig:evolsupp}
\end{figure}

\begin{table*}[h]
\centering
\begin{tabular}{l | c | c | c | c}
\toprule 
Inference Metrics& Time per iteration & Number of Iterations &  Total Time & Memory
 \\ & (sec/iter)  & & & (GB)\\
\midrule 
     Csgm-Langevin &  0.34 & 1155 
       & 6 min. 38 sec. & 5.3 \\ 
     Csgm-Langevin+ES  & 0.69 & 974
       & 11 min. 16 sec. & 5.3\\
     AM-Langevin & 0.80 & 1155
      & 15 min. 26 sec. & 5.3 \\
     SMRD &  1.14  & 902
      & 14 min. 54 sec. & 5.3 \\
\bottomrule
\end{tabular}

\caption{
Inference time and memory for different reconstruction methods.
}
\label{tbl:comparison_same_anatomy}
\end{table*}

\begin{table*}[h]
\centering
\begin{tabular}{l | c | c | c | c}
\toprule 
Inference Metrics&  Total Time & Number of Iterations\\
\midrule 
     $\sigma=0$ &   16 min. 47 sec. & 1007\\ 
    $\sigma=0.0025$ &  16 min. 22 sec. & 985\\
     $\sigma=0.005$ & 15  min. 38 sec. & 942\\
      $\sigma=0.0075$ & 14 min. 54 sec & 902  \\
\bottomrule
\end{tabular}

\caption{
Inference time for SMRD at different $\sigma$. For different noise levels, $T_{ES}$ for SMRD changes, reducing the total reconstruction time under high noise regimes.
}
\label{tbl:timing}
\end{table*}

\begin{figure}[h]
    \centering
    \includegraphics[width=1.0\textwidth,trim={0 2pt 0 0},clip]{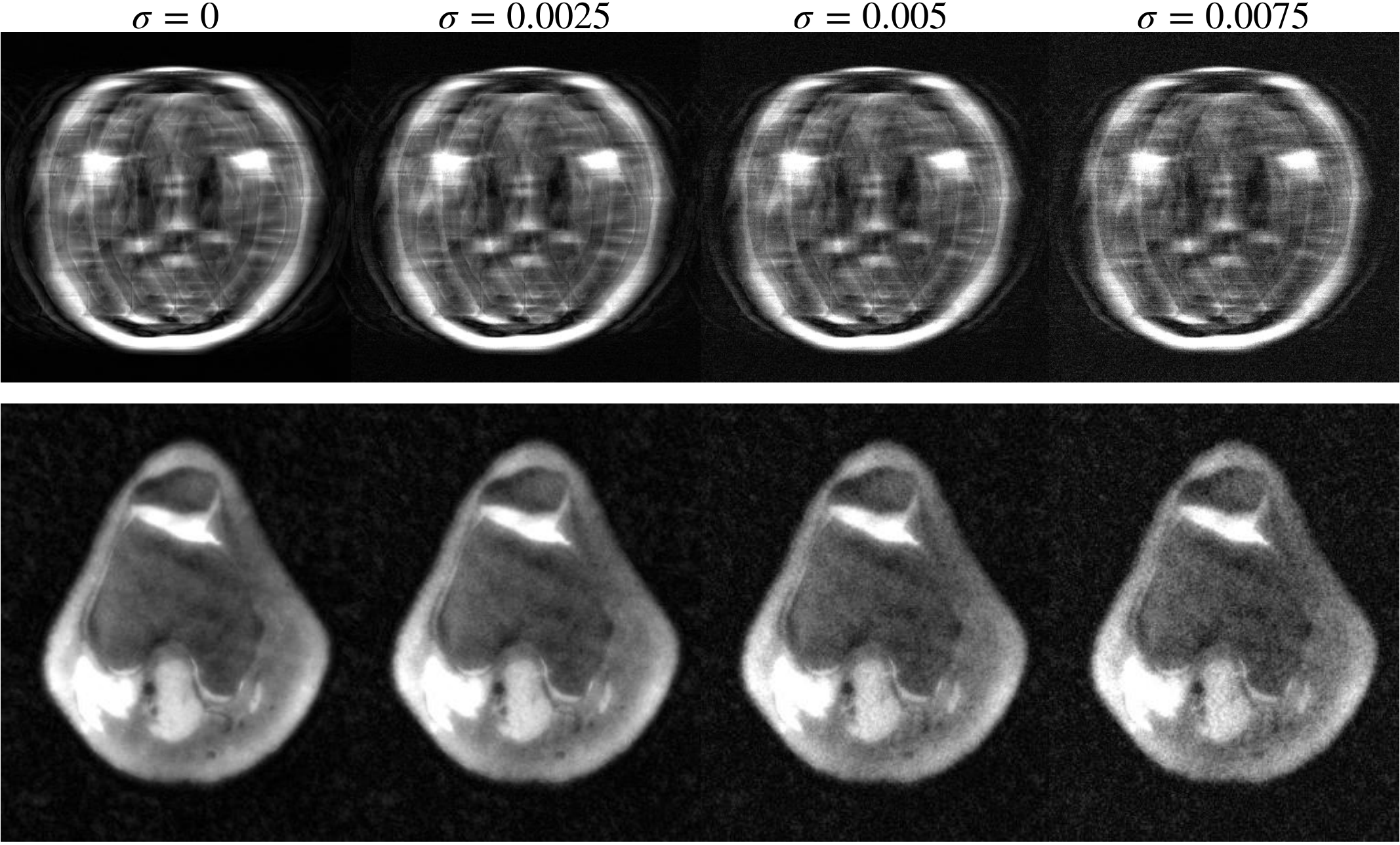}
    \caption{
    Example zero-filled images for different measurement noise levels for brain at $R =8$ (top) and for knee at $R = 12$ (bottom).
    }
    \label{fig:evol-supp}
\end{figure}

\end{document}